# 2x2 博弈实验中的社会旋动

Social Spiral Pattern in Experimental 2x2 Games

**许彬　　王志坚**[*1]
浙江大学实验社会科学实验室，杭州，310038

2010-09-18 Version (For Discussion)

## 摘要

演化博弈理论经常给出简洁的结果并用花蝴蝶般漂亮的图像展现其演化结果，2x2 博弈是其中的经典例子，其演化路径及其最终的均衡可以用一张图呈现。在实验经济学中，2x2 博弈实验的混合均衡就被认为是无规则的随机过程的总体分布。然而通过采用策略空间格点上的净流量矢量场分析方法，我们在 2x2 博弈实验数据中发现了净动态环流，也就是说，在格点的策略空间上，我们发现社会旋动的格局。该环流在格局上与演化博弈理论的结果十分相似。或者说在重复博弈中，人们不仅能够达到某种混合均衡状态，而且，更为重要的是，其达到均衡的动态格局恰恰是与演化理论预设的动态格局相一致。我们在已有的 2x2 实验数据中做出以上发现的。策略空间格点上的矢量场方法可以作为理论评估和实验设计的新方法。本文报告我们的发现：在 2x2 实验中社会存在着旋动的格局。

---

**Title:** Social Spiral in Experimental 2x2 Games

**Authors:** Bin Xu & Zhijian Wang* (Social Science Laboratory, Zhejiang University, Hangzhou, 310038, China)

**Abstract:** With evolutionary game theory, mathematicians, physicists and theoretical biologists usually show us beautiful figures of population dynamic patterns. 2x2 game (matching pennies game) is one of the classical cases. In this letter, we report our finding that, there exists a dynamical pattern, called as social spiral, in human subjects 2x2 experiment data(Goeree, Holt et al. 2003; Selten and Chmura 2008). In a flow/velocity vector field method, we explore the data in the discrete lattices of the macro-level social strategy space in the games, and then above spiral pattern emergent. This finding hints that, there exists a macro-level order beyond the stochastic process in micro-level. We notice that, the vector pattern provides an interesting way to conceal evolutionary game theory models and experimental economics data. This lattice vector field method provides a novel way for models evaluating and experiment designing.

---

*[1] Email: wangzj@zju.edu.cn　http://socexp.zju.edu.cn

# 介绍

# (Introduction)

近几十年来，2x2 模型在数学、物理学和理论生物学中得到广泛的研究(Hofbauer and Sigmund 2003; Becker, Chakrabarti et al. 2007; Szabo and Fath 2007)，在经济学中也一样(Friedman 1991)。并且，往往给出花蝴蝶般的理论上的演化轨迹图象。但是，在实证研究中(如实验经济学中)，人们认为满足图 1 条件的 2x2 博弈的混合均衡（Mixed equilibrium）是随机稳定的（stochastically stable），进而，采用策略空间分布均值与理论预言值的距离大小比较的方法，来判断不同理论的优劣(Foster and Young 1990; Goeree and Holt 1999; Goeree, Holt et al. 2003; Erev, Roth et al. 2007; Selten and Chmura 2008; Brunner, Camerer et al. 2009) 。

最近，Selten and Chmura(以下简称 SC)发表了一篇 2x2 博弈实验的文章（Selten and Chmura 2008），用实验结果与理论值的对比，评价并比较了几个领先的均衡模型(Selten and Chmura 2008)。该文报告的系列实验由 12 个 Game 组成，其中 game1~game6 为常数和博弈，game7~game12 为非常数和博弈，并且，它们分别依次是 game1~game6 的规律性变化，也就是说，12 个 game 中的常数和 game 与非常数和 game 配成 6 对，这一变换使得文章要比较的一些理论预言值在配对的博弈中保持不变，而另一些理论的预言值却发生变化，这样的设计可以多获得一种比较理论的途径。game1~game6 的每个 Game 都分别实施 12 个 session，game7~game12 中的每一个 game 分别实施了 6 个 session。每个 session 有 8 名参与者，分为横 player 和纵 player 两类各 4 名，横玩家和纵玩家随机配对博弈，每轮随机配对，共 200 轮。

实验的支付矩阵是 bi-matrix，如图 1 所示。实验结果用于比较几种不同的稳定概念。研究结果引起了凯姆勒等人的争论(Brunner, Camerer et al. 2009)以及 Selten 的再回应(Goerg 2010)。双方不仅对不同理论孰好孰坏有不同的判断，而且更重要的是引发了是否应该更多关注过程细节的不同观点。凯姆勒等人认为，正是因为引入了随机过程，才改进了理论（Camerer et al. 2009），而 Selten 等人则认为，我们首先应该关心的是静态稳定的状态，而不是陷入过程细节中。但无论哪一方，都没有对实验中被试策略选择的动态变化给出清晰而具体的图像，从而也就无从根据实验数据为他们的观点提供证据。我们认为，比理论优劣比较更为重要的是把实验过程中人的策略变化展示出来，观察分析其中的规律。之后才有可能对理论作出评价或对理论的改进提出建议。事实上，幸运的是，正如我们将要展示的那样，实验 2x2 博弈过程中，已经存在着有序的动态格局。就我们所知，迄今人们还未发现，实验室中的人群行为的总体演化路径呈现出整体旋转的动态特征。本文通过净流量矢量分析方法，首次展示 2x2 博弈中的社会宏观运动的旋动的特征。

| | |
|---|---|
| $a_{11}$<br>$b_{11}$ | $a_{12}$<br>$b_{12}$ |
| $a_{21}$<br>$b_{21}$ | $a_{22}$<br>$b_{22}$ |

图 1：2x2 Game 支付矩阵(Payoff Bi-Matrix)。混合均衡的条件是: $a_{11} > a_{21}, a_{22} > a_{12}, b_{12} > b_{11}$ & $b_{21} > b_{22}$，这个实验也叫 matching pennies game, cyclic 2x2 games 或 Buyer-Seller game.

# 净动态流矢量场

## (Definition: Net Vector Field of Social Motion on Lattices)

给定为数 $2N$ 的有限人群，2x2Game 在 up 的概率 p 和 left 的概率 q 构成的二维平面上形成 $(N+1)\times(N+1)$ 个格点的社会策略空间(Benaïm and Weibull 2003)。净动态流矢量表示的是社会策略空间的各个格点上的流量的净值，即流量的矢量和。这是一个宏观层次上的可测量。在重复博弈的任何一轮，系统必然处在其中的某一个格点 $\vec{r}=(i/N, j/N),(i,j\in[0,1,2,\ldots,N])$ 上（见图 2）。例如，在第 $t$ 轮和第 $t+1$ 轮，社会分别处于格点 $\vec{r}(t)=(i/N, j/N)$ 和 $\vec{r}(t+1)=(i'/N, j'/N)$ 上，那么，我们就说，在 t 时刻，系统的动态流矢量为 $\vec{f}(t)=\vec{r}(t+1)-\vec{r}(t)=(i'-i, j'-j)/N$。在多轮次的博弈中，策略空间上的每个格点都可能被多次经过，因此，我们可以根据式（1）对任一格点 $(i,j)$ 上在所有轮次中的 $\vec{f}$ 进行矢量求和得到该点的总的流量的矢量之和 $\vec{F}$（净动态流）：

$$\vec{F}(i,j)=\sum_{i'=0}^{N}\sum_{j'=0}^{N}\sum_{t=1}^{M}[\vec{r}(t+1)-\vec{r}(t)]\delta_{(i'-i_0)}\delta_{(j'-j_0)} \qquad (1)$$

其中 $\delta$ 表示的是狄拉克函数。各个格点的总净动态流矢量构成净动态流矢量场。

以 SC(2008)的实验为例，row players 和 column players 各 4 人，因此，社会共有 5 种 Up 的比率：0、0.25、0.5、0.75 和 1；同样，社会有 5 种 Left 的比率 0、0.25、0.5、0.75 和 1。因此，共有 25 个格点。图 2(a)给出了格点、矢量的示意图，而图 2(b)则根据定义，呈现了 SC 实验中的一个实验的实例，即 game5 的实验净动态流矢量图。可见，在各个格点的流矢量一起构成一个有序的图案。

依据上述计算得到的总净流矢量，我们还可以得到单位净流矢量及其图（见图 3）。单位流矢量是指每一个格点上的总净动态流矢量除以流矢量的发生次数，比如在某个格点经过 5 次，总净流矢量为（1，1）则单位流矢量为（0.2，0.2）。

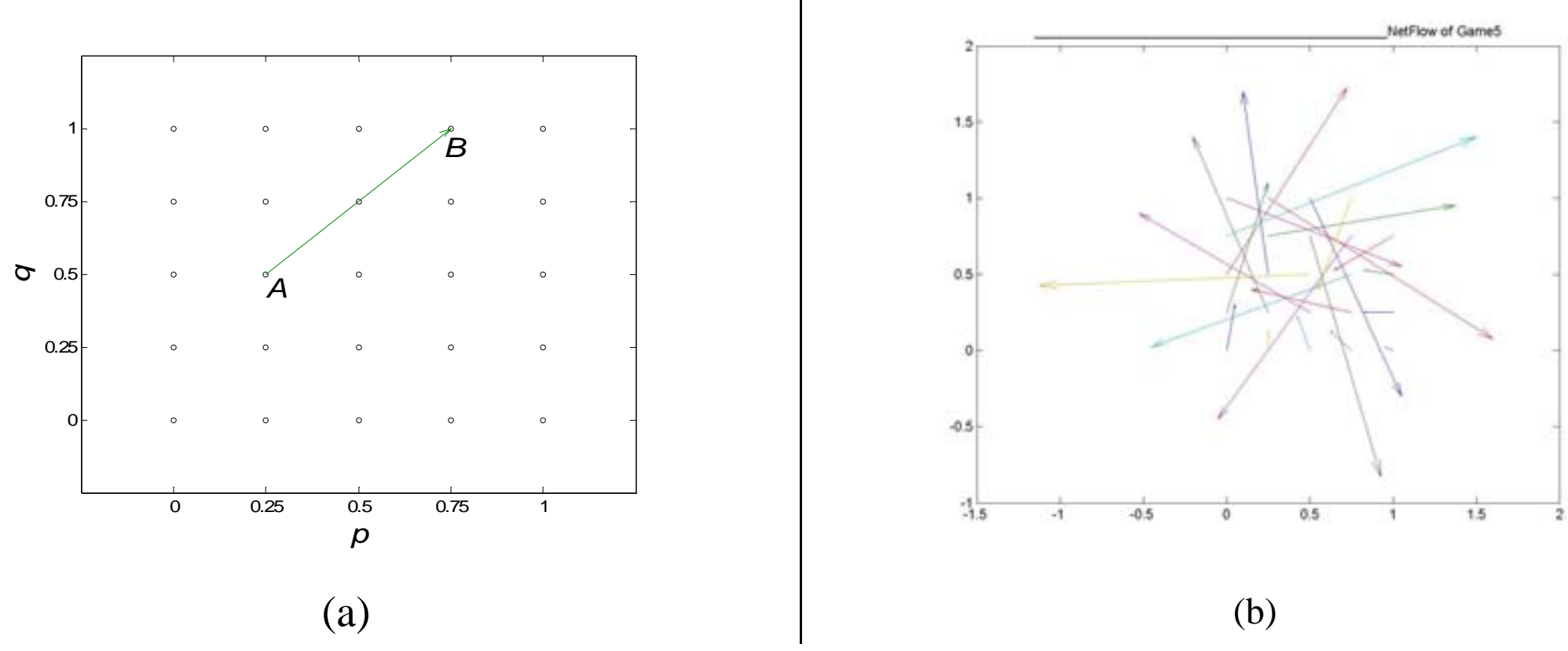

(a) (b)

**图 2 Definition Figure: Net Vector Field of Social Motion on Lattices 对于 2N 人参加的 2x2 实验， row player**

**有 N 人，所以，Up-Down 维度上的所有的可能的状态为 0, 1/N, 2/N,…,1，Left-Right 维度上也一样。图中以 SC(2008)的实验为例，row players 和 column players 各 4 人，社会共有 5 种 Up 的比率和 5 种 Left 的比率，策略空间共 25 个格点。任意一轮，群体状态必定落在这 25 个格点中的一个。图 2(a)是流矢量场示意图，图中横坐标 p 表示群体 up-down 中 up 的比率，纵坐标 q 表示 Left-Right 中 left 的比率，$\vec{r}(t)$ 为第 t 轮的状态，其落在格点 A(0.25, 0.5)上，表示 4 个 row player 中有 1 个选择 Up，在 4 个 column players 中有 2 个选择 Left。$\vec{r}(t+1)$ 为第 t+1 轮的状态，落在观点 B(0.75, 1)上，表示 4 个 row player 中有 3 个选择 Up，4 个 column players 中有 4 个选择 Left。带箭头的直线 AB 为净动态流矢量，表示第 t 轮人群系统在(0.25, 0.5)格点上的流量 $\vec{f}_{(t)}$ 为(0.5,0.5)。图 2(b)是净动态流矢量实例图，根据文中公式，对 SC2008 game5(Selten and Chmura 2008)的所有 12 个 session 的 2388 个流矢量求净值并以 40：1 的比例画出，也就是给出了 session 层面的平均径流矢量场图。**

# 社会旋动的实验证据

# (Social Spiral Pattern: Evident from Experimental Data)

由于 Selten 和 Chruma 的实验 (Selten and Chmura 2008)包含了 12 个 6 对 game 的实验，不仅有很大的实验规模，更由于其精心设计的实验参数使得其所包含的 6 对实验的预言值在策略空间上形成广泛的分布，不至于有偏向性上的嫌疑。因此，利用这一实验数据给出上述动态流矢量场图将是一个相对“中立”的选择。

在常数和博弈的 6 个 game 中，即，从 game1 到 game6，每个 game 有 12 个 Session，因此每个 game 有 2388 个观测值；在 6 个非常数和的 game 中，即从 game7 到 game12，每个 game 有 6 个 session。由于最后一轮不再有下一轮调整的机会，所以每个 session 的 200 轮中，只能统计到 199 个 $\vec{f}$；game1 到 game6 的每一个有 2388 个观测值，game7 到 game12 的每一个 game 有 1194 个观测值。我们根据本文上一部分的方法，分别计算每一个 game 的单位流矢量，并以 4：1 的比例给出 12 张图，见 Figure3。显然，12 张子图清楚地表明，12 个 game 无一例外地呈现出旋动的动态特征。

作为示意，图中红圈是实验均值观测点。另外一个例子是来自 Holt, Goeree 的数据(Goeree, Holt et al. 2003)，同样呈现出旋转的净流量场的特征。见 Append A2 可以注意到，SC 他们的实验观察值以及后来的辩论基于的观测(Brunner, Camerer et al. 2009; Goerg 2010)等于这 2400 个 $\vec{r} = (i/N, j/N)$ 观察值的平均值，而净动态流量场的方法可能是一种评估均衡点的一种新方法，见附录 A3。

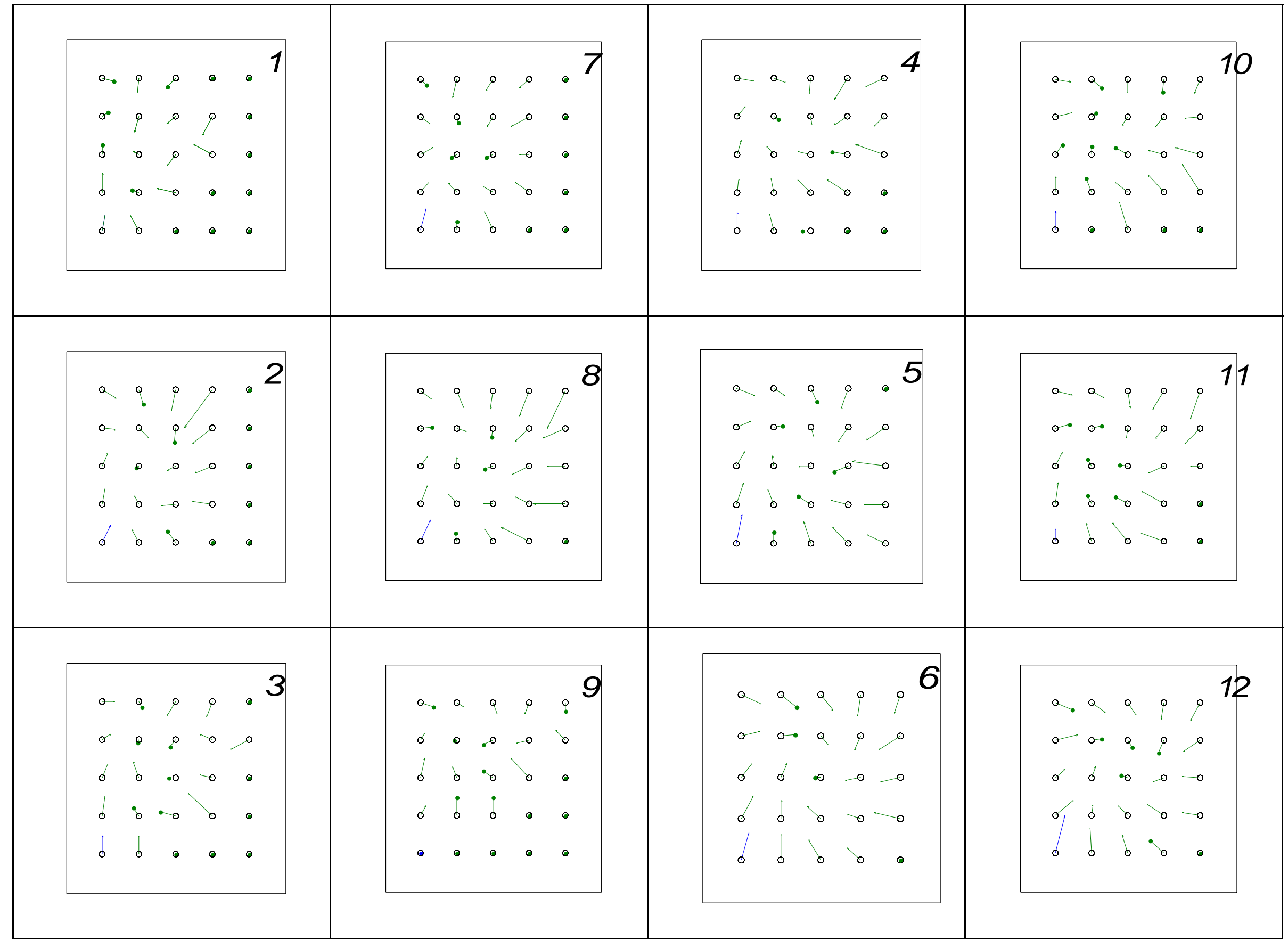

**图 3: 实验中的单位净动态流矢量图。图中的矢量长度是实际长度的 1/4。单位流矢量是总净流矢量除以流矢量发生的次数（流矢量不为 0 的次数）。各图右上角的数字表示的是对应 Game 的编码。该图说明，在 2x2 实验中，社会呈现旋转的动态特征。数据来源(Selten and Chmura 2008)**

# 演化理论相平面轨迹

# (Phase Plane of Evolution)

如此有规律的图像不仅使我们联想到演化理论的图像。那么，演化理论图像和真实实验博弈图像之间是一致的还是不一致的这一问题就被提了出来。本文这一部分试图对二者进行比较。

作为均衡理论，除了 Nash 理论还有 QRE(Quantal Response Equilibrium)(McKelvey and Palfrey 1995), IBE(Impulsive Balance Equilibrium)(Selten and Chmura 2008)等等。而有代表性的动态演化理论是最优反应 (Best response dynamics, BR)、分对数(Logit dynamics) 和 Maynard Smith 模型等等(Hofbauer and Sigmund 2003; Szabo and Fath 2007)。这些模型往往都做了无穷大人群和无限细分连续时间的近似，这样每一个 2x2 实验的社会演化动力学过程就可由微分方程组表达，其结果可以通过相平面图(phase portrait or phase plane) 方式呈现(Benaïm and Weibull 2003)。

这里，我们选 IBE Impulsive Balance Equilibrium）静态均衡理论和一个常用的演化动力学方程，最优反应动力学过程，展示全部 12 个 game 的理论图，更多的动力学和静态均衡的图形见附录（4）。最优反应过程动力学方程是：

$$\dot{\rho} = BR(\rho) - \rho \tag{2}$$

其中$\rho$是人群中的策略的比例矢量，BR 是最优响应函数(Szabo and Fath 2007)。

图 4 给出了 IBE 静态模型下 12 个 game 的 BR 演化图(各个 Game 的演化常微分方程组在图 4 的各个子图的顶端显示，带 IBE 理论预言值的图见附件)。对比图 3，我们发现，其理论上预言的演化轨迹与实际博弈中的旋转是一致的。在附件中，我们还给出 Logit 过程的 QRE 结果。直观可视地，我们判断，演化博弈论的演化方程可以解释实验中的动态轨迹。

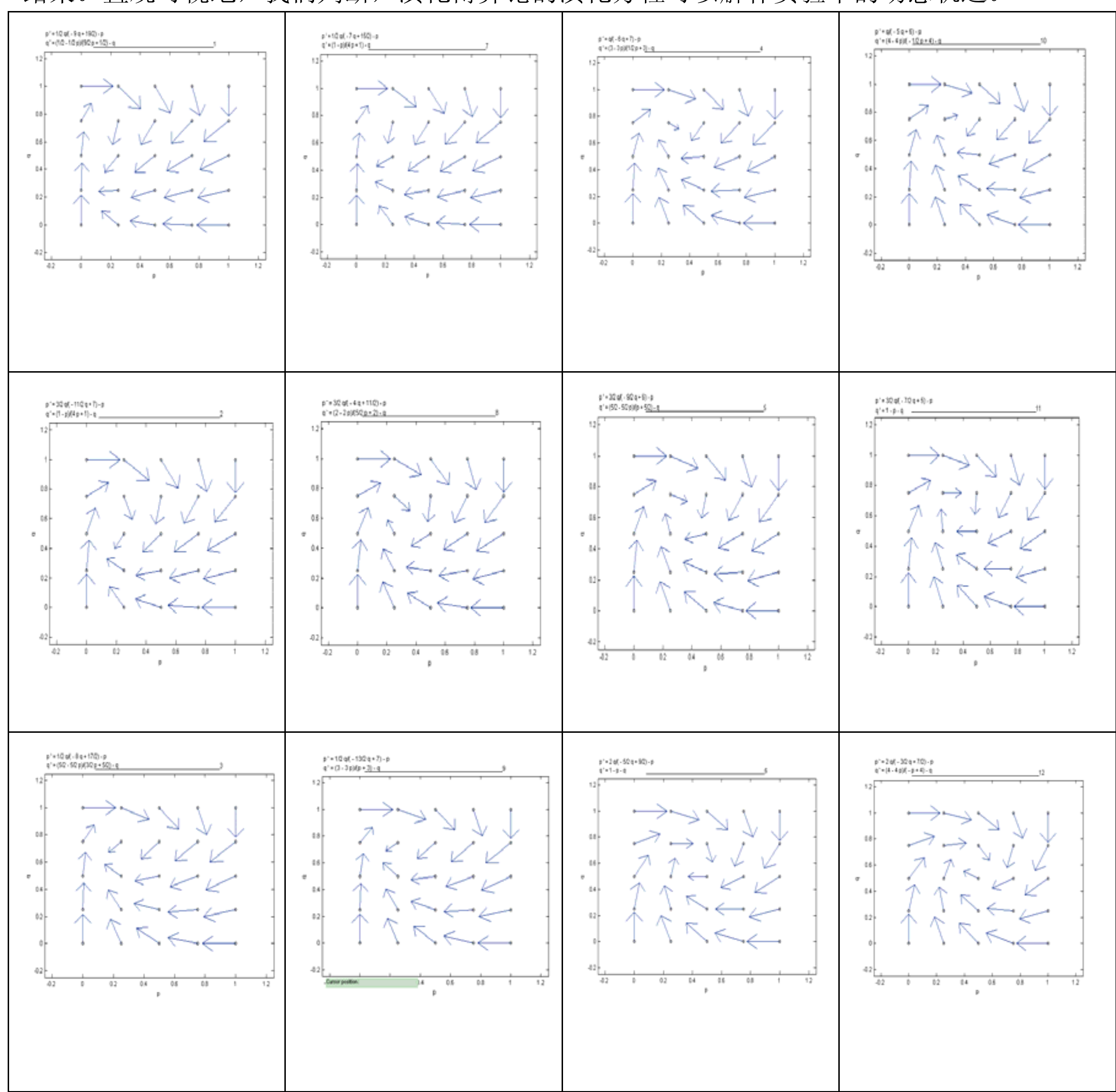

**图 4：12 个 game 的演化理论图。相应函数是 IBE, 而动力学方程为最优反应模式。各 Game 的微分方程见附件。QRE 对应的微分方程和相平面图结果分别见附件 A_QREEQ 和 A_QREPP。方法为 Domand Prince，步误差系数容忍度为 0.0001.**

# 总结

# (Summary)

演化理论在实验经济学中并非没有验证的先例，Van Huyck甚至提出了社会演化路径和演

化理论对应的矢量场进行对比的方法，他们将每五轮平均值投放在相空间中，发现实验(Van Huyck, Battalio et al. 1995)中群体在相空间中的移动轨迹与演化方程组给出的轨迹是大致是一致的(Van Huyck 2008，由于其为3x3实验结果，理论上和实验上都非旋转特征的格局。作为方法对比的例子，本文附录中只给出其8个Session的结果来说明格点矢量场方法的特征)。得到可视化的清晰的旋转格局图像，根据我们的知识，这是第一次，这里，我们采用的是格点上的净流量场分析方法

矢量场分析方法是力学中常见的问题，它可以得到关于相平面上矢量场的源、壑和旋转特征的数学描述。矢量场的应用，使得我们可以用矢量场的残值法( Residual Vector Field)对理论和实验数据进行比较,尤其通过对不同的理论与理论之间，不同的实验与实验之间，以及理论和实验之间的残值矢量场的格局的识别，我们可以得到一种基于几何的识别方法。另外一种方法是一对向量场二者间的投影的方法，也就是向量相似度法，形式化地表示如下：

$$D=\sum_{i,j}(1-v_{thor}\bullet v_{obs}/(|v_{thoe}\|v_{obs}|)) \qquad (3)$$

其中 ***D*** 表示二个向量场的差值的模,。***v*** 为单位净流量的理论或实验观测数值，求和是在实验观测数值不为 0 的各个点上进行求和的。这样，不同理论与实验的比较，可以通过 ***D*** 这个指标来标志。***D*** 数值小说明理论比较合理。再比如，可以用离散极坐标方法，计算旋转格局的下陷和旋转的程度。二维矢量场的识别的方法是天文学图像识别的常见的方法，而近年格局识别方法也在发展中(Bayly, KenKnight et al. 1998)，这一方法可以进一步发展成为检验理论的方法，这一工作将是作者另文的课题。

值得注意的是，本文分析的实验数据采自4对参与者的随机配对实验，也就是说，在每一个实验session中，只有4对8名参与者，但却得到了与有着大规模参与者和时间连续性假定的演化理论预言一致的旋转图像。这一结果令人惊讶，如同早期的只有4个买者和4个卖者的市场实验可以达到完全竞争的均衡一样。这前后两类只有8个人参加的实验都展示了社会层面的有序状态，而且与连续化的理论预言相近。这也是实验经济学的奇妙之处。

本文的贡献在于：（1）引进格点上的矢量分析方法对实验数据进行挖掘，首次呈现了社会旋转的整体图像。矢量分析方法对于把握社会运动规律提供了一种揭示动态过程特征的手段，该方法在实验经济学中的应用可以提供一种新的理论比较的路径；（2）采用可视化的方法，通过大参数范围的理论和实验图像的比较，展示演化博弈理论在格局上的可信性。